Unterrichtsmaterialien zur
ISS-Horizons-Mission von Alexander Gerst

In Kooperation mit DLR

Handreichung für Lehrpersonen:

# Wie fliegen Astronauten mit einer Rakete zur ISS?

**Klassen 7/8**

Markus Nielbock

22. Januar 2019

## Zusammenfassung

Diese Aktivität ermöglicht den Schülerinnen und Schülern nachzuempfinden, wie eine Rakete Besatzungen auf den Orbit der Internationalen Raumstation bringt. Da der Weg von den einfacheren Grundlagen hin zu einer realen Darstellung eines Raketenflugs recht komplex ist und Kenntnisse aus mehreren Klassenstufen benötigt, beschränkt sich die aktuelle Ausarbeitung auf die Einführung der Grundbegriffe und einfachen, idealisierten Anwendungen. Zentrale Bedeutung hat dabei die Raketengleichung. Weiterführende Ableitungen, die mehrstufige Raketen thematisieren, werden in einer gesonderten Ausarbeitung behandelt. Zur Einleitung in das Prinzip des Rückstoßes werden kurz einige Beispiele vorgestellt.

## Lernziele

Die Schülerinnen und Schüler

- machen sich mit dem Begriff des Impulses vertraut,
- erkennen, warum Raketen groß sein müssen,
- lösen eine Gleichung durch Iteration mittels Tabellenkalkulation,
- berechnen die Endgeschwindigkeit einer einstufigen Rakete.

## Materialien

- Arbeitsblätter (erhältlich unter: http://www.haus-der-astronomie.de/raum-fuer-bildung)
- Stift
- Taschenrechner
- Computer mit Tabellenkalkulation (z. B. MS Excel)

## Stichworte

Raumstation, ISS, Rakete, Rückstoß, Impuls, Schub, Dichte

## Dauer

90 Minuten

Unterrichtsmaterialien zur
ISS-Horizons-Mission von Alexander Gerst

In Kooperation mit

## Hintergrund

### Die Internationale Raumstation

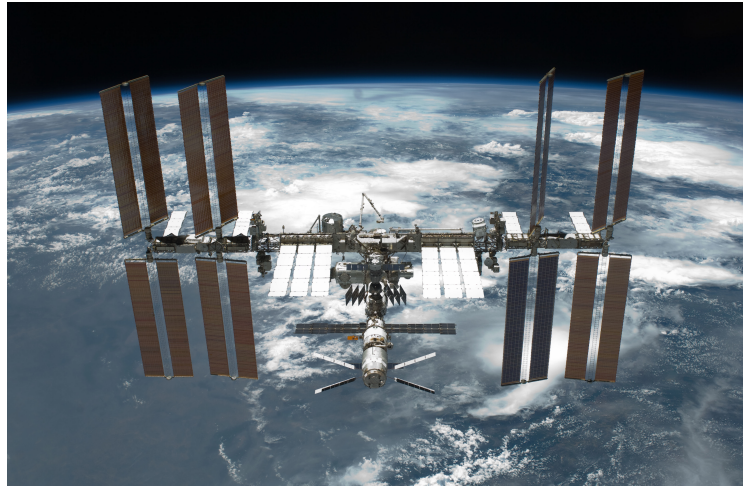

**Abbildung 1:** Die ISS im Jahre 2011 (Bild: NASA).

Seit 1998 wird die Internationale Raumstation (ISS, Abb. 1) aufgebaut (Loff 2015) und mittels einzelner Module (Abb. 2) ständig erweitert (Zak 2017a). Ihr Betrieb ist bis mindestens 2024 vorgesehen, wahrscheinlich aber sogar bis 2028 möglich (Sputnik 2016; Ulmer 2015). Die gesamte Struktur hat eine Masse von 420 t. Sie ist 109 m lang, 73 m breit (Garcia 2018b) und 45 m hoch (ESA 2014). Auf einer Bahnhöhe von etwa 400 km benötigt die ISS für eine Erdumrundung ungefähr 92 Minuten (Howell 2018a).

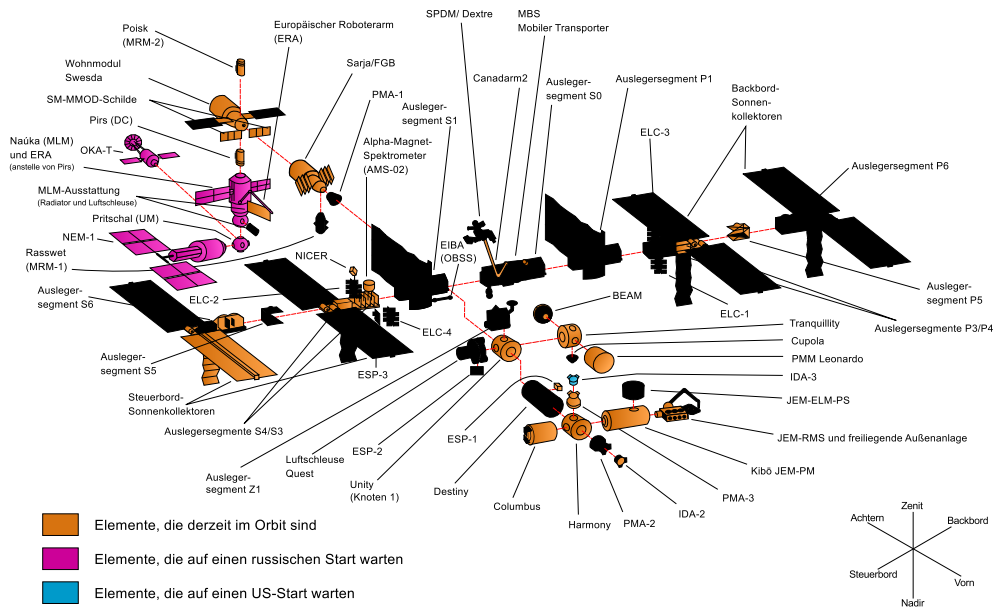

**Abbildung 2:** Die Module der ISS im Juni 2017 (Bild: NASA).

Die ISS ist ein internationales Projekt mit derzeit 15 beteiligten Nationen (ESA 2013b; Garcia 2018a). Sie dient als wissenschaftliches Forschungslabor für Fragestellungen, in denen z. B. der Einfluss der Gravitation auf der Erde bei Experimenten hinderlich ist. Es werden aber auch medizinische Themen der Astronautik behandelt, um langfristige Missionen innerhalb des Sonnensystems vorzubereiten.

**Sojus-FG-Rakete (Ракета «Союз-ФГ»)**

Seit dem letzten Start eines Space-Shuttles am 8. Juli 2011 sind die russischen Sojus-Raketen das einzige Trägersystem, das Menschen zur ISS bringen kann. Erst ab Frühjahr 2019 sind erste bemannte Testflüge mit US-amerikanischen Raumschiffen geplant (Malik 2018). Die Variante Sojus-FG (Abb. 3) wird dafür bereits seit 2001 routinemäßig benutzt. Alle Starts waren seitdem erfolgreich. Die Rakete basiert auf dem Modell R-7 (Howell 2018b; Zak 2018), das seit 1957 gebaut wurde. Sie war die Basis nahezu aller nachfolgenden Raketen, die in der Erkundung des Weltraums durch die damalige Sowjetunion und dem heutigen Russland entwickelt und genutzt wurden.

Die Sojus-FG besitzt je nach Nutzung drei bis vier Raketenstufen, wobei für den Transport des Sojus-Raumschiffs lediglich drei benötigt werden (Abb. 3). Das Raumfahrzeug hat einen eigenen Antrieb, der die Astronauten vom Orbit in 200 km Höhe auf 400 km bringt, wo die ISS sich befindet.

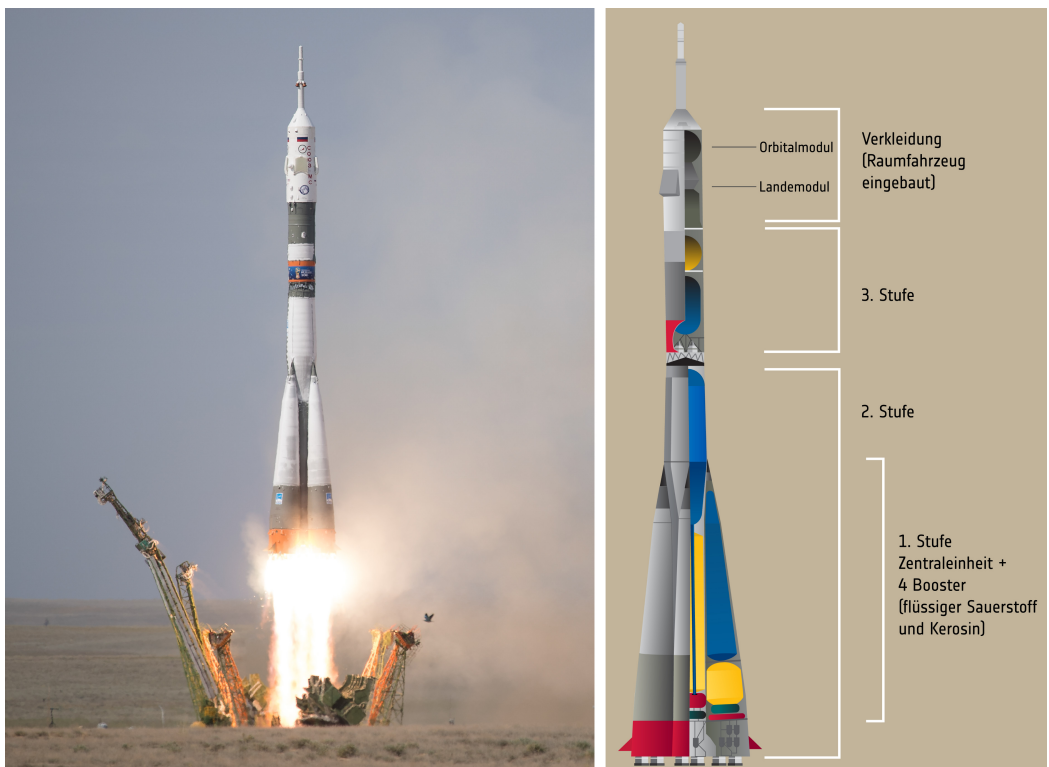

**Abbildung 3:** Links: Start der Sojus-FG-Rakete am 6. Juni 2018, mit der Alexander Gerst zur ISS flog (NASA/Joel Kowsky, https://www.flickr.com/photos/nasahqphoto/41898551494/ (CC BY-NC-ND 2.0). Rechts: Schematische Darstellung einer Sojus-FG-Rakete, https://www.flickr.com/photos/europeanspaceagency/35788717360/in/album-72157684209960351/ (ESA/Ausschnitt, deutsche Beschriftung: M. Nielbock).

In dieser Konfiguration besitzt die Rakete eine Höhe von 49,5 m (ESA 2013a) und eine Startmasse von etwa 310 t. Sie kann eine Nutzlast von bis zu 7,4 t in einen erdnahen Orbit bringen (Zak 2017b).

### Triebwerke und Stufen der Sojus-FG-Rakete

Die russische Sojus-FG-Rakete, mit der Alexander Gerst zur ISS flog, besteht aus drei Raketenstufen mit jeweils eigenen Triebwerken (NPO Energomash 2017; Zak 2017b). Sie alle werden mit einer raffinierten Form von Kerosin, genannt RD-1 (rocket propellant) (Leitenberger 2013), und flüssigem Sauerstoff (LOX, liquid oxygen) betrieben.

Die erste Stufe besteht aus vier identischen Triebwerken vom Typ RD-107A, die seitlich an der Rakete angebracht sind. Man nennt diese Konfiguration auch Booster. Die zweite Stufe ist die Zentraleinheit, die am Start zusammen mit der ersten Stufe gezündet wird. Das Triebwerk vom Typ RD-108A brennt jedoch länger als die erste Stufe. Nach dem Abwerfen der zweiten Stufe zündet die dritte Stufe, die mit einem Triebwerk RD-0110 ausgestattet ist. Sie hebt schließlich das Sojus-Raumschiff auf eine Höhe von etwa 200 km von wo es sich eigenständig an die ISS annähert.

### Schubkraft eines Raketentriebwerks

Wir kennen bereits das Prinzip des freien Falls. Ein Objekt mit einer Masse $m$ wird durch die Gravitationskraft der Erde angezogen. Diese Kraft führt dazu, dass dieses Objekt – einmal losgelassen – auf die Erde fällt. Dabei nimmt die Geschwindigkeit stetig zu. Die Rate, mit der die Geschwindigkeit sich ändert, nennt man Beschleunigung. Dieser Zusammenhang ist auch als die *Grundgleichung der Mechanik* oder *2. Newtonsches Axiom* bekannt. Daher kann man schreiben:

$$F_g = m \cdot a \qquad (1)$$

$$\text{mit: } a = \frac{\Delta v}{\Delta t} \qquad (2)$$

Diese Gleichung setzt also die auf $m$ wirkende Kraft $F_g$ in Beziehung zur Beschleunigung $a$, die es erfährt. In Bodennähe kann die Kraft vereinfacht durch $m \cdot g$ geschrieben werden, wobei $g$ die Erdbeschleunigung ist. Wir nehmen hier den Wert am Äquator der Erde an ($g = 9,8\,\text{m/s}^2$).

$$F_g = m \cdot g = m \cdot a \qquad (3)$$

Mit Gl. 2 erhält man dann:

$$m \cdot g = m \cdot \frac{\Delta v}{\Delta t} \qquad (4)$$

Die Geschwindigkeit der Masse $m$ nimmt daher im freien Fall je Zeiteinheit $\Delta t$ um $\Delta v$ zu, also in einer Sekunde um 9,8 m/s. Wir sehen hier auch, dass ohne weitere äußere Krafteinwirkung (z. B. Luftreibung) die Masse $m$ sich wegkürzt. Daraus folgt, dass der Geschwindigkeitszuwachs eines Objekts nicht von seiner Masse, sondern lediglich von der wirkenden Erdbeschleunigung $g$ abhängt.

Mit einer Rakete möchten wir das genaue Gegenteil erreichen, nämlich eine Nutzlast entgegen der wirkenden Gravitation nach oben befördern. Hierfür wird eine Kraft benötigt, die man als Schubkraft bezeichnet. Hierfür könnte man schlicht $F_S$ schreiben. Es hat sich jedoch die Schreibweise $S$ etabliert. Diese Schubkraft, oder kurz Schub, wird durch den Auswurf von verbranntem Treibstoff unter hoher Geschwindigkeit $w$ erzeugt. Man benutzt hier das Formelzeichen $w$, um die Geschwindigkeit der Triebwerkgase von der Geschwindigkeit der Rakete zu unterscheiden. Die Masse des Treibstoffs wird der Gesamtmasse der Rakete entzogen und wird deshalb mit $\Delta m$ bezeichnet.

Der Treibstoffdurchsatz $\mu = \Delta m/\Delta t$ zeigt also an, mit welcher Rate Treibstoff verbraucht wird und sich die Masse der Rakete ändert. Insgesamt erhält man:

$$S = \frac{\Delta m}{\Delta t} \cdot w = \mu \cdot w \qquad (5)$$

Die Einheit des Schubs entspricht der einer Kraft, somit: $\text{kg} \cdot \text{m/s}^2 = \text{N}$. Um mit der Rakete abheben zu können, muss also vom Betrag her stets $S > F_g$ gelten, wobei die Masse der Rakete $m_R$ ständig pro $\Delta t$ um $\Delta m$ abnimmt. Die Fähigkeit einer Rakete, den Erdboden zu verlassen, hängt also von der Startmasse der Rakete, dem Treibstoffdurchsatz $\mu$ und der Ausströmgeschwindigkeit $w$ ab. Die letzteren beiden Größen sind charakteristisch für die verschiedenen Triebwerke, die in der Raumfahrt benutzt werden.

**Impuls und spezifischer Impuls**

Rechnerisch ist der Impuls $p$ nichts weiter als das Produkt aus Masse und Geschwindigkeit.

$$p = m \cdot v \qquad (6)$$

Physikalisch beschreibt der Impuls $p$ den Bewegungszustand eines Objekts. Jede Änderung des Impulses $\Delta p$ kann nur durch die Aufwendung einer Kraft erfolgen. Je länger sie wirkt, desto stärker verändert sich der Impuls. Umgekehrt geht die Ausübung einer Kraft durch ein sich bewegendes Objekt mit einer Änderung des Impulses einher. Es gilt daher:

$$\Delta p = F \cdot \Delta t \Leftrightarrow F = \frac{\Delta p}{\Delta t} \qquad (7)$$

In der Raketentechnik hat sich der Begriff des *spezifischen Impulses* $I_{\text{sp}}$ eingebürgert. Hierunter versteht man den Impuls pro Massenelement des von einem Triebwerk ausgestoßenen Verbrennungsprodukts. Definiert ist er als das Produkt des über die Brenndauer $\tau$ gemittelten Schubs $S$ geteilt durch die Masse des verbrannten Treibstoffs. Das Produkt aus dem Schub und der Brenndauer ist jedoch nichts anderes als der zeitlich gemittelte Impuls der ausströmenden Gases.

$$I_{\text{sp}} = \frac{\bar{S} \cdot \tau}{m_{\text{treib}}} \qquad (8)$$

$$= \frac{\bar{p}}{m_{\text{treib}}} \qquad (9)$$

Die Einheit von $I_{\text{sp}}$ ist somit m/s, also eine Geschwindigkeit. $I_{\text{sp}}$ ist daher im wesentlichen nichts anderes als die Ausströmgeschwindigkeit $w$, die sich jedoch in der Praxis zeitlich ändert. Das erkennt man auch durch den Vergleich mit der Definition des Schubs in Gl. 5. Sowohl der Schub als auch der spezifische Impuls sind von der äußeren Druckbedingungen abhängig, da das Triebwerk gegen diesen Druck arbeitet. Deswegen steigen typischerweise $S$, $I_{\text{sp}}$ und $w$ mit zunehmender Höhe und abnehmendem atmosphärischem Druck.

Oft bezieht man den spezifischen Impuls nicht auf die Masse des Gases, sondern auf ihr Gewicht unter der Einwirkung der Erdbeschleunigung $g$ (Messerschmid und Fasoulas 2011). Die resultierende Einheit entspricht dann derjenigen der Zeit. Wir werden jedoch stets mit der Definition nach Gl. 8 arbeiten.

$$I_{\text{sp}}^{\star} = \frac{\bar{S} \cdot \tau}{m_{\text{treib}} \cdot g} \qquad (10)$$

**Raketengleichung**

Die sogenannte Raketengleichung oder auch Ziolkowski-Gleichung – benannt nach dem russischen Weltraumpionier Konstantin Ziolkowski, der diese Grundgleichung 1903 aufstellte – beschreibt die Bewegung einer einstufigen, kräftefreien Rakete. Sie lässt sich über zwei äquivalente Prinzipien herleiten – die Grundgleichung der Mechanik bzw. das 2. Newtonschen Axiom und die Impulserhaltung.

Aus der Grundgleichung der Mechanik lässt sich eine Bewegungsgleichung für das Verhalten einer Rakete mit dem Schub $S$ nach Gl. 5 aufstellen.

$$F = m \cdot a \Leftrightarrow S = m_R \cdot a_R \tag{11}$$

In diesem Fall ist $m_R$ von der Zeit $t$ abhängig, denn mit der Zündung der Rakete verbrennt Treibstoff mit dem Durchsatz $\mu = \Delta m/\Delta t$. Wenn wir annehmen, dass $S$ konstant ist (was in der Realität nicht stimmt) nimmt die Beschleunigung im Laufe der Zeit also zu. Somit kann man schreiben:

$$\mu \cdot w = (m_0 - \mu \cdot t) \cdot a_R(t) \tag{12}$$

Dabei ist $m_0$ die Masse der Rakete bei Brennbeginn. Für kleine Veränderungen von $\Delta t$ können wir näherungsweise schreiben:

$$\frac{\Delta m}{\Delta t} \cdot w = (m_0 - \Delta m) \cdot \frac{\Delta v_R}{\Delta t} \tag{13}$$

$$\Leftrightarrow \Delta m \cdot w = (m_0 - \Delta m) \cdot \Delta v_R \tag{14}$$

$$\Leftrightarrow \Delta v_R = \frac{\Delta m}{m_0 - \Delta m} \cdot w \tag{15}$$

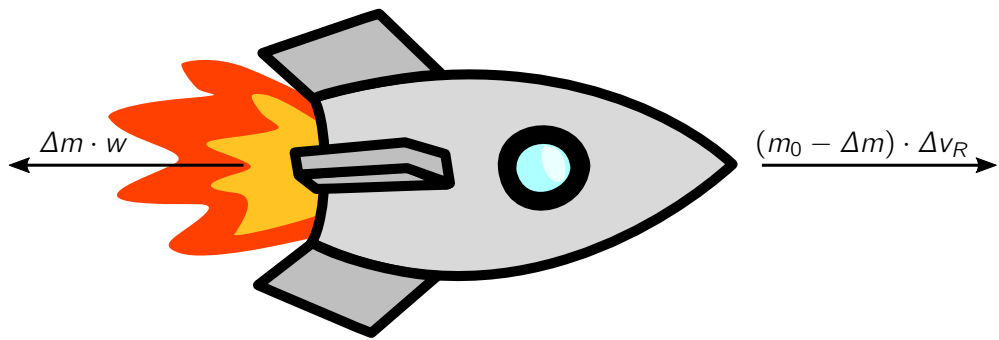

In Gl. 14 erkennt man die Impulserhaltung mit den Teilimpulsen des ausströmenden Gases und des Raketenkörpers. Für genügend kleine $\Delta m = \mu \cdot \Delta t$ kann man mit Hilfe von Gl. 15 die Geschwindigkeitsänderung der Rakete für bekannte Werte von $\mu$ und $w$ berechnen. Das ist besonders hilfreich wenn man die Endgeschwindigkeit einer Raketenstufe mit der Methode der kleinen Schritte annähern möchte.

Tatsächlich muss man für das korrekte Ergebnis ein Integral lösen, wobei man von Gl. 12 ausgehend $v(t) = \int a(t)\,dt$ auswertet. Wir nehmen an, dass $\mu$ und $w$ konstant sind.

$$v_R(t) = \int_{t_0}^{t_1} a(t)\,dt = \int_{t_0}^{t_1} \frac{\mu \cdot w}{m_0 - \mu \cdot t}\,dt = -w \cdot \int_{t_0}^{t_1} \frac{-\mu}{-\mu \cdot t + m_0}\,dt \tag{16}$$

Mit $\frac{d}{dx}\ln(a \cdot x + b) = \frac{a}{a \cdot x + b}$ folgt:

$$[v_R(t)]_{t_0}^{t_1} = -w \cdot [\ln(m_0 - \mu \cdot t)]_{t_0}^{t_1} \quad (17)$$

$$\Leftrightarrow v_R(t_1) - v_R(t_0) = \Delta v_R = -w [\ln(m_0 - \mu \cdot t_1) - \ln(m_0 - \mu \cdot t_0)] \quad (18)$$

$$\Leftrightarrow \Delta v_R \stackrel{\substack{t_0 = 0 \\ \mu \cdot t_1 = m_{\text{treib}}}}{=} -w \cdot [\ln(m_0 - m_{\text{treib}}) - \ln(m_0)] \quad (19)$$

$$\Leftrightarrow \Delta v_R \stackrel{m_B = m_0 - m_{\text{treib}}}{=} -w \cdot \ln\left(\frac{m_B}{m_0}\right) \quad (20)$$

$$\Leftrightarrow \Delta v_R = w \cdot \ln\left(\frac{m_0}{m_B}\right) \quad (21)$$

Hier bedeuten $m_0$ die Gesamtmasse der Rakete zu Brennbeginn und $m_B$ die Masse der Rakete bei Brennende. Daraus ersieht man, dass eine Änderung der Geschwindigkeit vom natürlichen Logarithmus des Massenverhältnisses zu Beginn und Ende der Triebwerkzündung abhängt. Damit muss für eine bestimmte Geschwindigkeitsänderung eine überproportional zusätzlich große Treibstoffmenge umgesetzt werden, was zu einer deutlich größeren Rakete mit einer deutlich höheren Masse führt. So benötigt man bei einer Rakete mit einem Massenverhältnis von Treibstoff und leerer Raketenstufe von vier für eine Verdopplung der Endgeschwindigkeit sechs mal so viel Treibstoff.

# Aktivität: Raketenflug

## Vorbereitung für Lehrpersonen

Lesen Sie das Kapitel mit den Hintergrundinformationen sorgfältig. Zusätzliche Literatur finden Sie am Ende dieses Dokuments.

Die Schülerinnen und Schüler benötigen für diese Aktivität bereits Vorkenntnisse zu dem Grundlagen der Mechanik, d. h. Grundgleichung der Mechanik, Begriff des Impulses, Begriff der Kraft.

Machen Sie sich mit den Aufgaben der Schülerinnen und Schüler vertraut. Fertigen Sie ausreichend Kopien der Arbeitsblätter an. Für die Aktivität benötigen sie Taschenrechner sowie Laptops/PCs, auf denen MS Excel vorinstalliert ist. Es wäre hilfreich, wenn die Schülerinnen und Schüler bereits erste Erfahrungen mit Tabellenkalkulationen hätten.

Dieses Material beinhaltet wegen des großen Umfangs des Themengebiets „Raketenflug" lediglich eine Einführung in die Grundbegriffe der Raketenantriebstechnik sowie eine erste Berechnung einer Endgeschwindigkeit nach dem Umsetzen des Treibstoffs in kinetische Energie. Hierzu werden vereinfachte Annahmen gemacht. Äußere Kräfte werden in dieser Ausarbeitung nicht berücksichtigt. Eine Vertiefung, die auch den Einfluss von mehreren Raketenstufen sowie den Beitrag der Gravitation berücksichtigt, werden in einer nachfolgenden Aktivität behandelt.

## Thematische Einführung (Vorschlag)

Zur Einführung in das Thema bietet es sich an, die Schülerinnen und Schüler zu ihren bisherigen Kenntnissen und Erfahrungen zur Raumfahrt zu befragen. Fragen Sie sie, ob sie einen (deutschen) Astronauten kennen und was sie über ihn wissen. Als thematische Einführung können zudem Fotos und kurze Videos von der Internationalen Raumstation gezeigt werden. So gibt es einen faszinierenden Livestream von der ISS.

https://youtu.be/gnKt_QHehZc

Fragen Sie die Schülerinnen und Schüler, wie die Astronauten und Kosmonauten zur ISS kommen. Sobald richtigerweise Raketen genannt werden, versuchen Sie ermitteln, ob sie wissen, wie eine Rakete funktioniert.

Schließlich wird das Prinzip des Rückstoßes von entscheidender Bedeutung sein. Daher bietet es sich an, Beispiele für die Wirkung des Rückstoßprinzips zu behandeln, die entweder bekannt sind oder Interesse wecken.

Ein recht einfaches Beispiel stellt ein Luftballon dar, aus dem Luft entweicht und so durch den Raum fliegt. Abhängig vom Alter oder der Vorbildung der Schülerinnen und Schüler kann die Lehrperson Luftballons verteilen und zum Experimentieren animieren. Lassen Sie die Schülerinnen und Schüler diskutieren, was den Luftballon fliegen lässt.

Alexander Gerst demonstriert in einem Experiment auf der ISS das Prinzip des Rückstoßes einer Rakete.

https://youtu.be/bkdXM0L6ghI

Um selbst aktiv zu werden, können die Lernenden zudem zum Bau von Fahrzeugen motiviert werden, die auf das Rückstoßprinzip zurückgreifen. Interessante Varianten führen den bereits erwähnten Ballon mittels eines angeklebten Strohhalms entlang einer Schnur oder bewegen durch Aufkleben ein Auto.

https://www.kids-and-science.de/experimente-fuer-kinder/detailansicht/datum/2009/11/22/die-ballonrakete-an-der-schnur.html

http://www.zauberhafte-physik.net/raketenautos-22/

Unter folgender Internetadresse findet sich eine Anleitung zum Bau eines Propellerboots.

https://tuduu.org/projekt/flinke-propellerboote

Für einen direkten Zugang zur Raumfahrt und der Nutzung von Raketen können die Schülerinnen und Schüler die Internetseite des Portals DLR_next studieren. Hier finden sie viele Informationen und Anregungen.

https://www.dlr.de/next/desktopdefault.aspx/tabid-6283/

Mit etwas technischem Geschick können die Lernenden selbst eine Wasserrakete bauen. Dies kann beispielsweise mit dem Fach Technik verknüpft werden oder im Rahmen einer AG als Gruppenprojekt durchgeführt werden. Es existiert eine Vielzahl von Bauanleitungen mit unterschiedlichen Schwierigkeitsgraden.

https://www.dlr.de/next/desktopdefault.aspx/tabid-6297/
http://www.raketfuedrockets.com

Einige Aspekte des Flugs mit der Sojus-Rakete zur ISS sind in dem folgenden Video zusammengefasst.

https://www.dlr.de/next/desktopdefault.aspx/tabid-12713

Leiten Sie abschließend gemeinsam mit den Lernenden die vereinfachte Raketengleichung (Gl.15) gemäß der Darstellung auf Seite 6 her.

## Schub und Treibstoff

### 1. Schubkraft eines Kugelstoßers

Ein Kugelstoßer der Weltklasse beschleunigt die 7,257 kg schwere Kugel innerhalb von einer Sekunde auf eine Abwurfgeschwindigkeit von 14 m/s. Berechne den dadurch auf den Kugelstoßer ausgeübten Schub.

### 2. Schubkraft eines Raketentriebwerks

Die russische Sojus-FG-Rakete, mit der Alexander Gerst zur ISS flog, besteht aus drei Raketenstufen mit jeweils eigenen Triebwerken. Die erste Stufe besteht aus vier identischen Triebwerken vom Typ RD-107A, die seitlich an der Rakete angebracht sind. Man nennt diese Konfiguration auch Booster. Beim Start erzeugen sie einen Schub von insgesamt $S = 4146400\,\text{N}$ und die Ausströmgeschwindigkeit beträgt $w = 2580\,\text{m/s}$ (Zak 2017b). Berechne den Treibstoffdurchsatz $\mu$.

### 3. Treibstoffmenge und Tanks einer Rakete

Die zweite Raketenstufe der Sojus-FG besitzt ein Triebwerk des Typs RD-108A. Als Treibstoff dient ein Gemisch aus Kerosin und flüssigem Sauerstoff (LOX, liquid oxygen), das im Volumenverhältnis 1:2,43 einen Schub von $S = 792650\,\text{N}$ erzeugt. Die Ausströmgeschwindigkeit beträgt während der Brenndauer von 280 s im Mittel $w = 2525\,\text{m/s}$. Berechne $\mu$.

Berechne nun, wie viel Kerosin und Sauerstoff (in kg) während der Brenndauer umgesetzt werden. Wie hoch ist die Gesamtmasse der Treibstoffe?

Wir wollen nun berechnen, wie groß die Treibstofftanks für Kerosin und Sauerstoff sein müssen. Dazu solltest du dir in Erinnerung rufen, was die Dichte eines Stoffs, hier einer Flüssigkeit, bedeutet und wie man sie berechnet. Wir können annehmen, dass die Dichten der beiden Stoffe $\varrho_{\text{Ker}} = 830\,\text{kg/m}^3$ bzw. $\varrho_{\text{LOX}} = 1144\,\text{kg/m}^3$ betragen (Uffrecht und Poppe 2002, S. 86). Mit dieser Information berechnest du nun die Volumina, die das Kerosin und das LOX einnehmen.

Nun sollst du ausrechnen, wie hoch die jeweiligen Tanks der Raketenstufe sein müssen. Da wir für die Tanks eine Zylinderform annehmen, musst du dafür wissen, wie man das Volumen eines Zylinders berechnet. Aufgrund des Durchmessers der Rakete ist der Durchmesser der Tanks auf jeweils 2,2 m beschränkt. Wie hoch müssten die Tanks sein?

Wie steht das in Relation zur Höhe der Raketenstufe, die 27,1 m misst?

## Endgeschwindigkeit einer einstufigen Rakete

Betrachte den Idealfall einer kräftefreien, einstufigen Rakete. Das bedeutet, wir nehmen an, dass

1. die Rakete von äußeren Kräften wie die Schwerkraft oder die atmosphärische Reibung nicht beeinflusst wird,
2. die Rakete den kompletten Treibstoff in einem Brennvorgang umsetzt.

Die Rakete habe die folgenden Eigenschaften.

| Eigenschaft | Formelzeichen | Wert |
|---|---|---|
| Gesamtmasse Rakete | $m_0$ | 320 t |
| Leermasse Rakete | $m_B$ | 20 t |
| spezifischer Impuls | $w$ | 2800 m/s |
| Brenndauer | $\tau$ | 600 s |

Berechne hieraus die Masse des Treibstoffs $m_\text{treib}$, den Treibstoffdurchsatz $\mu$ und den Schub $S$.

## 1. Genäherte Lösung durch Methode der kleinen Schritte

Wegen der Beschleunigung der Rakete nimmt ihre Geschwindigkeit ständig zu.

**Frage:** Warum kann man mit Gl. 15 die Endgeschwindigkeit nicht korrekt ausrechnen?

Wir können diese Gleichung dennoch nutzen, wenn wir die Brenndauer der Rakete in kleinere Einzelschritte unterteilen. Damit kannst du die Endgeschwindigkeit der Rakete sehr gut abschätzen. Dieses Prinzip soll nun genauer untersucht werden. Erzeuge Dir dafür zunächst drei Tabellen, die folgende Spalten enthalten:

| Schritt | Raketenmasse $m_0$ (kg) | Masse bei Brennende $m_B$ (kg) | Geschwindigkeitszuwachs $\Delta v_R$ (m/s) | Endgeschwindigkeit $v_R$ (m/s) |
|---|---|---|---|---|
| 0 | 320000 | – | 0 | 0 |

Schritt 0 bedeutet, dass wir hier die Startparameter eintragen. Die Rakete bewegt sich noch nicht. Deswegen kann man auch noch keine Masse bei Brennende eintragen.

Füge zur ersten Tabelle eine Zeile hinzu, zur zweiten 3 und zur dritten 5. In die erste Spalte „Schritte" trägst du jeweils fortlaufende Zahlen ein, so dass diese Spalte die Zeilen durchnummeriert. Dies sind die jeweiligen Einzelschritte, die wir oben bereits erwähnt haben.

Berechne nun für jede Tabelle das $\Delta m$, also die Masse, die pro Rechenschritt vom Triebwerk ausgestoßen wird. Hierzu teilst du ganz einfach den am Anfang vorhandenen Treibstoff in so viele Pakete auf, wie die jeweilige Tabelle Schritte aufweist. Tabelle 1 besitzt nur einen Schritt. Somit gilt dort:

$$\Delta m = \frac{m_\text{treib}}{\text{Anzahl der Schritte}} = \frac{m_0 - m_B}{1} = 300\,\text{t} = 300000\,\text{kg} \tag{22}$$

Nun kannst du die Werte für Schritt 1 ausrechnen. Die aktuelle Raketenmasse in der Tabelle ist immer die Masse am Brennende des vorherigen Schritts. Da in Tabelle 1 Schritt 0 nur den Anfangszustand wiedergibt, hat sich die Masse nicht geändert. Berechne nun die Masse bei Brennende $m_B$. Bedenke, dass bei jedem Schritt $\Delta m$ des Treibstoffs verbrannt werden.

**Frage:** Warum fällt die Restmasse der Rakete nicht auf Null ab?

Mithilfe von Gl. 15 kannst du nun die Geschwindigkeitsänderung $\Delta v_R$ der Rakete ausrechnen, die während dieses Schritts zur vorherigen Geschwindigkeit hinzu kommt. Die Endgeschwindigkeit am Ende eines Berechnungsschritts erhältst du, wenn du $\Delta v_R$ zu $v_R$ aus dem vorherigen Schritt hinzurechnest. In Tabelle 1 ist $\Delta v_R = v_R$, da $v_R$ zuvor Null betrug.

Füge diese Rechenoperationen nun für die beiden anderen Tabellen durch. Beachte, dass dort $\Delta m$ andere Werte annimmt. Berechne jeden Schritt einzeln.

**Frage:** Wie unterscheiden sich die Geschwindigkeiten der Rakete am Ende? Diskutiere den Unterschied mit deinen Mitschülern und versuche, eine Erklärung zu erhalten.

**Frage:** Was passiert mit der Endgeschwindigkeit der Rakete, wenn du die Anzahl der Schritte erhöhst?

### 2. Methode der kleinen Schritte mit MS Excel

Erzeuge nun solch eine Tabelle in MS Excel und führe die Berechnungen mit Operationen der Tabellenkalkulation durch. Wähle hierfür eine deutlich größere Anzahl von Schritten. Um die Ergebnisse zu vergleichen, sprich dich mit deinen Nachbarn bei der Wahl der Anzahl der Schritte ab, z. B. 30, 60 und 150.

**Frage:** Was stellst du bezüglich den Endgeschwindigkeiten fest, wenn du dein Ergebnis mit deinen Mitschülern vergleichst? Erkläre die Beobachtung.

Recherchiere, mit welcher Geschwindigkeit die ISS um die Erde fliegt. Wie passt dieser Wert zur Geschwindigkeit der Rakete?

### Lernkontrolle und Abschluss

Neben den korrekten Antworten und Ergebnissen, die von einzelnen Schülerinnen und Schülern erzielt wurden, kann eine weitere Lernkontrolle spielerisch erreicht werden. Hier bietet sich beispielsweise *Kahoot!* an. Dies ist eine interaktive Online-Plattform für verschiedene didaktische Methoden. Typischerweise definiert die Lehrperson zunächst ein Quiz, das dann von den Schülerinnen und Schülern mit ihren Smartphones gespielt wird. Als Spielleiter können Sie die Anzahl der korrekten Antworten pro Frage einsehen und grafisch für alle darstellen. Zudem besteht die Möglichkeit, dass Sie nachsehen können, wer richtige und falsche Antworten gegeben hat und darauf gezielt eingehen. So können Sie z. B. nachfragen, warum jemand eine bestimmte Antwort ausgewählt hat.

Ein solches Quiz ist bereits vorbereitet und kann unter folgendem Link abgerufen werden.

https://play.kahoot.it/#/k/78434d13-69fd-4d5f-88f4-0233123e9cd6

# Lösungen

## Schub und Treibstoff

### 1. Schubkraft eines Kugelstoßers

$$\begin{aligned} S &= \mu \cdot w \\ &= \frac{\Delta m}{\Delta t} w \\ &= \frac{7{,}257\,\text{kg}}{1\,\text{s}} \cdot 14\,\text{m/s} \\ &= 101{,}6\,\frac{\text{kg\,m}}{\text{s}^2} \end{aligned}$$

### 2. Schubkraft eines Raketentriebwerks

$$\begin{aligned} S &= \mu \cdot w \\ \Leftrightarrow \mu &= \frac{S}{w} \\ &= \frac{4146400\,\text{N}}{2580\,\text{m/s}} \\ &= 1607{,}1\,\frac{\text{kg}}{\text{s}} \end{aligned}$$

Das Triebwerk stößt pro Sekunde 1607,1 kg Treibstoff aus.

### 3. Treibstoffmenge und Tanks einer Rakete

$$\begin{aligned} \mu &= \frac{S}{w} \\ &= \frac{792650\,\text{N}}{2525\,\text{m/s}} \\ &= 313{,}9\,\frac{\text{kg}}{\text{s}} \end{aligned}$$

Die Gesamtmasse berechnet man durch die Multiplikation des Treibstoffdurchsatzes $\mu$ mit der Brenndauer $\tau$.

$$m_{\text{treib}} = \mu \cdot \tau = 313{,}9\,\frac{\text{kg}}{\text{s}} \cdot 280\,\text{s} = 87897{,}8\,\text{kg} = 87{,}9\,\text{t}$$

Die Massenanteile von Kerosin und LOX addieren sich zur gesamten Treibstoffmasse: $m_{\text{treib}} = m_{\text{Ker}} + m_{\text{LOX}}$. Das Mischungsverhältnis von Kerosin und LOX ist wie folgt gegeben.

$$\frac{V_{\text{Ker}}}{V_{\text{LOX}}} = \frac{1}{2{,}43}$$

Mit

$$V = \frac{m}{\varrho}$$

folgt:

$$\begin{aligned}
\frac{V_{\text{Ker}}}{V_{\text{LOX}}} &= \frac{1}{2,43} \\
\Leftrightarrow \frac{m_{\text{Ker}} \cdot \varrho_{\text{LOX}}}{m_{\text{LOX}} \cdot \varrho_{\text{Ker}}} &= \frac{1}{2,43} \\
\Leftrightarrow \frac{m_{\text{Ker}}}{m_{\text{LOX}}} &= \frac{\varrho_{\text{Ker}}}{\varrho_{\text{LOX}}} \cdot \frac{1}{2,43} \\
\Leftrightarrow m_{\text{Ker}} &= \frac{\varrho_{\text{Ker}}}{\varrho_{\text{LOX}}} \cdot \frac{m_{\text{LOX}}}{2,43}
\end{aligned}$$

Somit folgt:

$$\begin{aligned}
m_{\text{treib}} &= m_{\text{LOX}} + m_{\text{Ker}} = m_{\text{LOX}} + \frac{\varrho_{\text{Ker}}}{\varrho_{\text{LOX}}} \cdot \frac{m_{\text{LOX}}}{2,43} = m_{\text{LOX}} \cdot \left(1 + \frac{\varrho_{\text{Ker}}}{\varrho_{\text{LOX}}} \cdot \frac{1}{2,43}\right) \\
\Leftrightarrow m_{\text{LOX}} &= \frac{m_{\text{treib}}}{\left(1 + \frac{\varrho_{\text{Ker}}}{\varrho_{\text{LOX}}} \cdot \frac{1}{2,43}\right)}
\end{aligned}$$

Man erhält mit $\varrho_{\text{Ker}} = 830\,\text{kg/m}^3$ und $\varrho_{\text{LOX}} = 1144\,\text{kg/m}^3$:

$$\begin{aligned}
m_{\text{LOX}} &= \frac{87897,8\,\text{kg}}{\left(1 + \frac{830\,\text{kg/m}^3}{1144\,\text{kg/m}^3} \cdot \frac{1}{2,43}\right)} \\
&= \frac{87897,8\,\text{kg}}{1,299} = 67688,2\,\text{kg} = 67,7\,\text{t}
\end{aligned}$$

$$m_{\text{Ker}} = m_{\text{treib}} - m_{\text{LOX}} = 87897,8\,\text{kg} - 67688,2\,\text{kg} = 20209,6\,\text{kg} = 20,2\,\text{t}$$

Somit erhält man mit $\varrho_{\text{Ker}} = 830\,\text{kg/m}^3$ und $\varrho_{\text{LOX}} = 1144\,\text{kg/m}^3$ die Volumina von Kerosin und LOX.

$$\begin{aligned}
V_{\text{Ker}} &= \frac{m_{\text{Ker}}}{\varrho_{\text{Ker}}} = \frac{20209,6\,\text{kg}}{830\,\text{kg/m}^3} = 24,349\,\text{m}^3 \\
V_{\text{LOX}} &= \frac{m_{\text{LOX}}}{\varrho_{\text{LOX}}} = \frac{62271,6\,\text{kg}}{1144\,\text{kg/m}^3} = 59,168\,\text{m}^3
\end{aligned}$$

**Hinweis!** Man kann die Rechnung auch umkehren, indem man zunächst die Volumina ausrechnet und dann über die Dichten die Massen ermittelt.

Als Treibstofftanks nehmen wir Zylinder mit dem Radius $r = 1,1\,\text{m}$ an. Mit der Höhe $h$ kann der Volumeninhalt eines Zylinders wie folgt berechnet werden:

$$V_{\text{Zyl}} = \pi \cdot r^2 \cdot h$$

$$\Leftrightarrow h = \frac{V_{\text{Zyl}}}{\pi \cdot r^2}$$

Die Höhen der beiden Tanks ermitteln sich dann zu:

$$h_{\text{Ker}} = \frac{V_{\text{Ker}}}{\pi \cdot r^2} = \frac{24{,}349\,\text{m}^3}{\pi \cdot (1{,}1\,\text{m})^2} = 6{,}4\,\text{m}$$

$$h_{\text{LOX}} = \frac{V_{\text{LOX}}}{\pi \cdot r^2} = \frac{59{,}168\,\text{m}^3}{\pi \cdot (1{,}1\,\text{m})^2} = 15{,}6\,\text{m}$$

Die beiden Tanks weisen eine Gesamthöhe von 22,0 m auf. Etwa 5 m entfallen somit auf die Verbindungskupplungen zwischen den Raketenstufen und auf das Triebwerk.

**Endgeschwindigkeit einer einstufigen Rakete**

$$m_{\text{treib}} = m_0 - m_B = 320\,\text{t} - 20\,\text{t} = 300\,\text{t}$$

$$\mu = \frac{m_{\text{treib}}}{\tau} = \frac{300\,\text{t}}{600\,\text{s}} = 0{,}5\,\frac{\text{t}}{\text{s}} = 500\,\frac{\text{kg}}{\text{s}}$$

$$S = \mu \cdot w = 500\,\frac{\text{kg}}{\text{s}} \cdot 2800\,\frac{\text{m}}{\text{s}} = 1400000\,\frac{\text{kg}\,\text{m}}{\text{s}^2}$$

**1. Genäherte Lösung durch Methode der kleinen Schritte**

**Frage:** Warum kann man mit Gl. 15 die Endgeschwindigkeit nicht korrekt ausrechnen?

Die Gleichung lautet:

$$\Delta v_R = \frac{\Delta m}{m_0 - \Delta m} \cdot w$$

Da wir hier den Unterschied der Raketenmasse zwischen Zündung und Brennende betrachten, kann die ermittelte Geschwindigkeit nur ein Mittelwert sein. Der ist nur dann richtig, wenn sich die Geschwindigkeit während der Brenndauer nicht ändert. Das ist aber ausdrücklich nicht der Fall, denn die Rakete beschleunigt während der gesamten Brennphase.

Wir können die Gleichung jedoch für eine schrittweise Näherung der tatsächlichen Endgeschwindigkeit nutzen, wenn $\Delta m$ klein genug ist und der Brennvorgang in entsprechend viele Teilschritte aufgetrennt wird.

**Tabelle 1**

$$\Delta m = \frac{m_{\text{treib}}}{\text{Anzahl der Schritte}} = \frac{300000\,\text{kg}}{1} = 300000\,\text{kg}$$

| Schritt | Raketenmasse $m_0$ (kg) | Masse bei Brennende $m_B$ (kg) | Geschwindigkeitszuwachs $\Delta v_R$ (m/s) | Endgeschwindigkeit $v_R$ (m/s) |
|---|---|---|---|---|
| 0 | 320000 | – | 0 | 0 |
| 1 | 320000 | 20000 | 42000 | 42000 |

**Frage:** Warum fällt die Restmasse der Rakete nicht auf Null ab?

Die Restmasse ist die Leermasse der Rakete. Die bleibt am Ende des Brennvorgangs übrig.

**Tabelle 2**

$$\Delta m = \frac{m_{\text{treib}}}{\text{Anzahl der Schritte}} = \frac{300000\,\text{kg}}{3} = 100000\,\text{kg}$$

| Schritt | Raketenmasse $m_0$ (kg) | Masse bei Brennende $m_B$ (kg) | Geschwindigkeitszuwachs $\Delta v_R$ (m/s) | Endgeschwindigkeit $v_R$ (m/s) |
|---|---|---|---|---|
| 0 | 320000 | – | 0 | 0 |
| 1 | 320000 | 220000 | 1273 | 1273 |
| 2 | 220000 | 120000 | 2333 | 3606 |
| 3 | 120000 | 20000 | 14000 | 17606 |

**Tabelle 3**

$$\Delta m = \frac{m_{\text{treib}}}{\text{Anzahl der Schritte}} = \frac{300000\,\text{kg}}{5} = 60000\,\text{kg}$$

| Schritt | Raketenmasse $m_0$ (kg) | Masse bei Brennende $m_B$ (kg) | Geschwindigkeitszuwachs $\Delta v_R$ (m/s) | Endgeschwindigkeit $v_R$ (m/s) |
|---|---|---|---|---|
| 0 | 320000 | – | 0 | 0 |
| 1 | 320000 | 260000 | 646 | 646 |
| 2 | 260000 | 200000 | 840 | 1486 |
| 3 | 200000 | 140000 | 1200 | 2686 |
| 4 | 140000 | 80000 | 2100 | 4786 |
| 5 | 80000 | 20000 | 8400 | 13186 |

**Frage:** Wie unterscheiden sich die Geschwindigkeiten der Rakete am Ende? Diskutiere den Unterschied mit deinen Mitschülern und versuche, eine Erklärung zu erhalten.

Die Geschwindigkeit nimmt mit zunehmender Anzahl der Schritte ab. Die Anzahl der Schritte steht für ein Aufteilen der aufgrund der Beschleunigung der Rakete steigenden Geschwindigkeit. Die ist anfangs geringer und nimmt langsam zu. Mit nur einem Schritt wird der anfänglich geringe Geschwindigkeitszuwachs stark überschätzt.

**Frage:** Was passiert mit der Endgeschwindigkeit der Rakete, wenn du die Anzahl der Schritte erhöhst?

Die Unterschiede der verschiedenen Ansätze nehmen mit der Anzahl der Rechenschritte ab. Es ist zu erwarten, dass mit zunehmender Anzahl der Schritte sich der Wert einem festen Betrag annähert.

## 2. Methode der kleinen Schritte mit MS Excel

Die Lösungen mit 30, 60, 150 Schritten sehen wie folgt aus.

**Tabelle 1:** Endgeschwindigkeit der in der Aufgabe spezifizierten Rakete durch iterative Rekursion nach 30, 60 und 150 Schritten.

| Schritte | Endgeschwindigkeit (m/s) |
|---|---|
| 30 | 8476 |
| 60 | 8106 |
| 150 | 7897 |

Als Beispiel für eine solche Tabelle ist in Abb. 4 eine Ausführung mit 30 Iterationsschritten dargestellt.

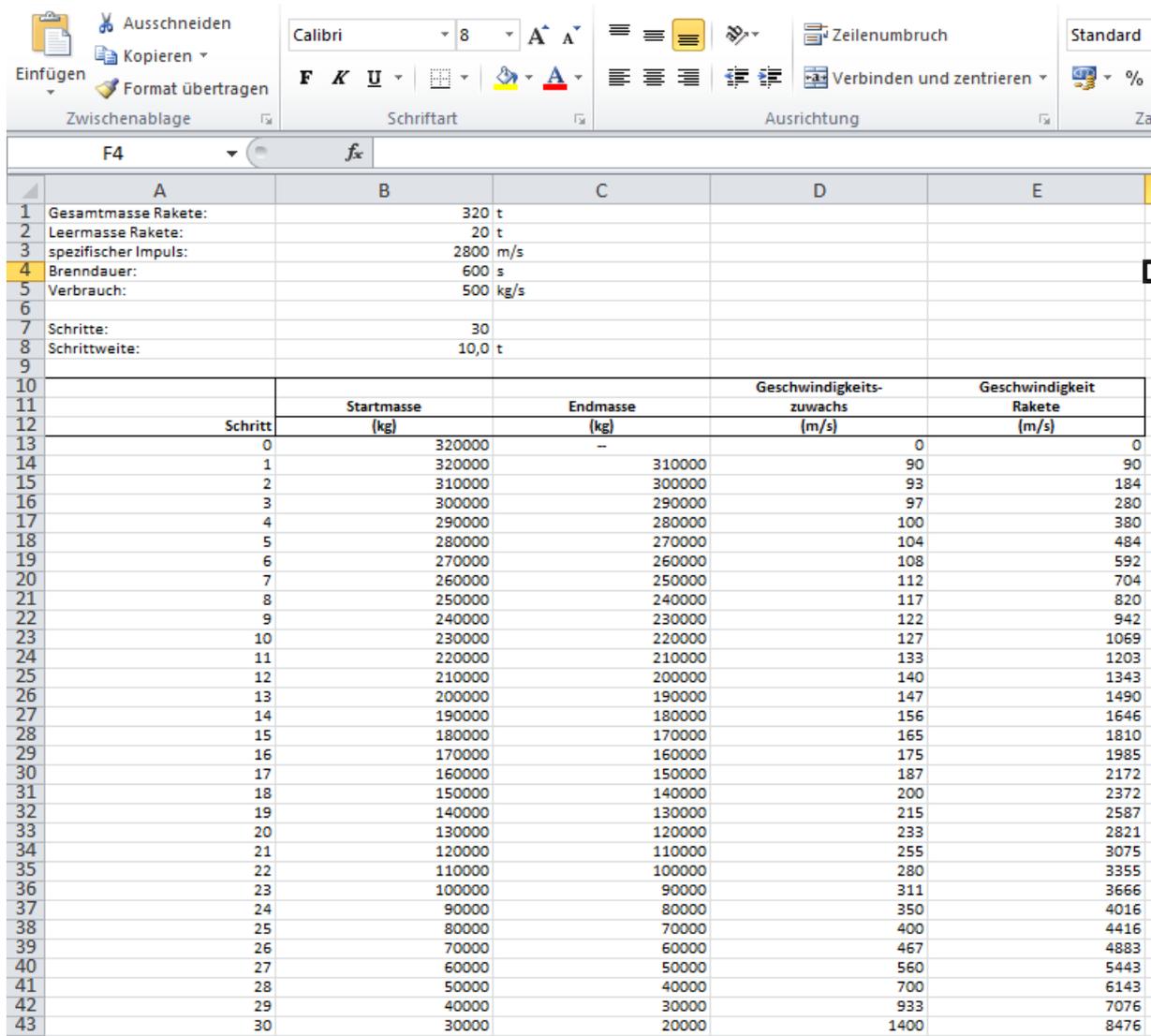

**Abbildung 4:** Eine mit MS Excel erstellte Tabelle, die mit 30 Schritten iterativ die Endgeschwindigkeit der Rakete ermittelt.

**Frage:** Wie unterscheiden sich die Geschwindigkeiten der Rakete am Ende? Diskutiere den Unterschied mit deinen Mitschülern und versuche, eine Erklärung zu erhalten.

Im Vergleich mit den Lösungen wird deutlich, dass die Unterschiede mit der zunehmenden Anzahl der Schritte immer geringer werden.

**Frage:** Was passiert mit der Endgeschwindigkeit der Rakete, wenn du die Anzahl der Schritte erhöhst?

Sie streben gegen einen Wert, der der exakten Lösung entspricht. Mit Gl. 21 erhält man 7763 m/s.

Die Geschwindigkeit der ISS schwankt leicht. Sie liegt jedoch üblicherweise in einem Bereich um 7,7 km/s. Die aktuelle Bahngeschwindigkeit wird hier dargestellt:

https://www.n2yo.com/?s=25544

Damit würde die Rakete in etwa die Geschwindigkeit der ISS erreichen.

## Danksagung

## Literatur

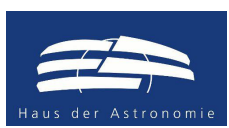
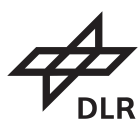
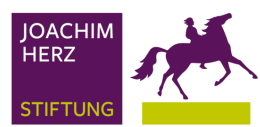